\documentclass[%
 aip,
 amsmath,amssymb,
 reprint,%
]{revtex4-2}

\usepackage{graphicx}

\usepackage{dcolumn}
\usepackage{bm}
\usepackage{xcolor}
\usepackage{bbm}

\usepackage[utf8]{inputenc}
\usepackage[T1]{fontenc}
\usepackage{mathptmx}
\usepackage[hidelinks]{hyperref}
\begin{document}

\title[Enhanced absorption in magnetized plasmas]{Enhanced collisionless laser absorption in strongly magnetized plasmas}

\author{Lili Manzo}
\affiliation{%
Lawrence Livermore National Laboratory, Livermore, California 94551, USA
}%
\author{Matthew R. Edwards}%
\affiliation{%
Lawrence Livermore National Laboratory, Livermore, California 94551, USA
}%
\affiliation{Stanford University, Stanford, California 94305, USA}
\author{Yuan Shi}
\email{shi9@llnl.gov}
\affiliation{%
Lawrence Livermore National Laboratory, Livermore, California 94551, USA
}%

\date{\today}

\begin{abstract}
Strongly magnetizing a plasma adds a range of waves that do not exist in unmagnetized plasmas and enlarges the laser-plasma interaction (LPI) landscape. 
In this paper, we use particle-in-cell (PIC) simulations to investigate strongly magnetized LPI in one dimension under conditions relevant for magneto-inertial fusion experiments, focusing on a regime where the electron-cyclotron frequency is greater than the plasma frequency and the magnetic field is at an oblique angle with respect to the wave vectors. 
We show that when electron-cyclotron-like hybrid wave frequency is about half the laser frequency, the laser light resonantly decays to magnetized plasma waves via primary and secondary instabilities with large growth rates. These distinct magnetic-field-controlled instabilities, which we collectively call two-magnon decays, are analogous to two-plasmon decays in unmagnetized plasmas.
Since additional phase mixing mechanisms are introduced by the oblique magnetic field, collisionless damping of large-amplitude magnetized waves substantially broadens the electron distribution function, especially along the direction of the magnetic field.
During this process, energy is transferred efficiently from the laser to plasma waves and then to electrons, leading to a large overall absorptivity when strong resonances are present.
The enhanced laser energy absorption may explain hotter-than-expected temperatures observed in magnetized laser implosion experiments and may also be exploited to develop more efficient laser-driven x-ray sources.

\end{abstract}

\maketitle

\section{\label{sec:intro}Introduction}
In both direct-drive and indirect-drive approaches to inertial confinement fusion (ICF), laser-plasma interactions (LPI) control the coupling and symmetry of implosions.\cite{Kirkwood2013,myatt2014multiple} The addition of a background magnetic field, which intends to improve particle and heat confinement,\cite{Chang11,davies2017laser,walsh2021magnetized}may strongly affects LPI physics.

Previous research on magnetized LPI has found increased second harmonic generation,\cite{jha2007} altered laser wakefield acceleration,\cite{krasovitskii2004} suppression of the modulational instability,\cite{jha2005,jia2017kinetic} reduced Raman scattering,\cite{Liu2018}  and the presence of additional parametric instabilities.\cite{edwards2019laser,los2021magnetized}
In ICF relevant regimes, previous particle-in-cell (PIC) simulations on weakly magnetized LPI, where the electron cyclotron frequency $\omega_c$ is much smaller than the plasma frequency $\omega_p$, have shown that magnetization increases the amount of energy absorbed by raising the threshold for kinetic inflation and decreasing backscattering.\cite{winjum2018mitigation,zhou2021suppression}
Although the initial stage of magnetized implosions is in this weakly magnetized regime, where tens of tesla fields are applied by external coils, the implosions likely encounter stronger fields at later stages due to flux compression. With a convergence ratio of ten, the late-time magnetic fields may reach kilo-tesla level, \cite{knauer2010fluxcompression,sio2021diagnosing} where magnetization can affect LPI in very different ways. \cite{shi2018laser}

In this paper, we use PIC simulations to investigate kinetic effects when a strong background magnetic field $\mathbf{B}_0$ is present ($\omega_c > \omega_p$). 
Instabilities in this regime have been studied in the context of magnetic confinement when waves propagate nearly perpendicular to the magnetic field. \cite{porkolab1988parametric} In particular, it is well known that strong absorption occurs when the pump frequency $\omega\approx2\omega_c$. 
However, in inertial confinement, lasers usually propagate at oblique angles, so conditions for strong absorption are different. The absorption near second cyclotron harmonics is distinct from the absorption due to two-plasmon decay, which occurs when $\omega\approx2\omega_p$ in unmagnetized plasmas. \cite{Goldman1966parametric,Baldis1983growth,Regan2010suprathermal,Turnbull20}
Both absorption mechanisms are modified when waves propagate obliquely in magnetized plasmas, where the Langmuir and cyclotron waves are no longer linear eigenmodes. Instead, they hybridize to form a Langmuir-like eigenmode, which we call the P wave, and an electron-cyclotron-like eigenmode, which we call the F wave. 
In terms of these eigenmodes, two-plasmon decay occurs when $\omega$ is about twice the P wave frequency, and the decay when $\omega$ is about twice the F wave frequency shall be called two-magnon decay. The latter is the focus of this paper, and we will show that two-magnon decays are in fact comprised of a number of distinct mechanisms.

From our simulations, at least four distinct scenarios of two-magnon decays can be identified. First, in the reference scenario (SB0), where no F wave is strongly excited, the interaction is dominated by backscattering via the P wave. Second, in the primary-backward scenario (SB1), backscattering via the F wave becomes a dominant resonance. Third, in the primary-forward scenario (SF1), forward scattering via the F wave becomes dominant. Finally, in the secondary-forward scenario (SF2), backscattered light pumps strong F-wave mediated forward scattering. 
In Sec.~\ref{sec:plasmawaves}, we show how the spectra of waves that are spontaneously excited by the laser depend on $B_0$. 
In Sec.~\ref{sec:vdfs}, we show that plasma waves heat electrons and lead to anisotropic broadening of the distribution function. 
We inspect the energy balance of the system in Sec.~\ref{sec:energy} and show that in resonant cases there is an increase in plasma energy and a reduction in field energy, confirming that laser energy is absorbed via wave-particle interactions.

\section{\label{sec:params}Methods}
We use the EPOCH PIC code \cite{arber2015contemporary} to carry out fully kinetic relativistic simulations in one spatial dimension with three velocity components. 
The simulation parameters are chosen to be typical of inertial fusion experiments. A hydrogen plasma (ion mass $m_i = 1836 m_e$, where $m_e$ is the electron mass) with initial electron and ion temperatures $T_e = T_i = 1 \textrm{ keV}$ and initial electron density $n_e = 10^{20} \textrm{ cm}^{-3}$ extends uniformly over $100$ $\mu$m in $x$, with 25-$\mu$m-wide vacuum regions on both sides. 
The initial velocity distribution function $f(\mathbf{v})$ is Gaussian and symmetric in all velocity directions. 
A laser with vacuum wavelength $\lambda_0 = 0.351 \;\mu\textrm{m}$, intensity $I_0 = 1.6\times10^{16}\;\textrm{ W/cm}^2$, and linear polarization in the $y$ direction enters from the $-x$ boundary. Upon entering the magnetized plasma, the laser polarization axis rotates due to the Faraday effect and an oscillating ellipticity is induced due to birefringence. The incident laser intensity remains constant throughout the simulation, barring an initial sinusoidal ramp with temporal profile $I(t<T)=I_0\sin \frac{\pi t}{2T}$, where $T$ is set to 220 laser periods to reduce numerical artifacts. 
All simulations use transmissive boundary conditions for fields and particles, and a resolution of 40 cells per laser wavelength and 100 particles per cell. 
A uniform magnetic field $B_0$ is applied in the $x$-$z$ plane, making a $60^\circ$ angle with $x$, the direction of laser propagation. Such an oblique magnetic field is common for magnetized laser-driven implosion experiments. \cite{davies2017} Moreover, an oblique angle allows for general laser-plasma instabilities to occur, many of which have zero growth rate at exact parallel or perpendicular angles. \cite{shi2018laser}
For results reported in this paper, all parameters are kept constant except for $B_0$, which is scanned between 0 and \mbox{20 kT}. At even larger field strengths, the electron cyclotron frequency approaches the laser frequency, leading to strong instabilities that have been explored previously. \cite{edwards2019laser}

\begin{figure}
	\includegraphics[width=0.35\textwidth]{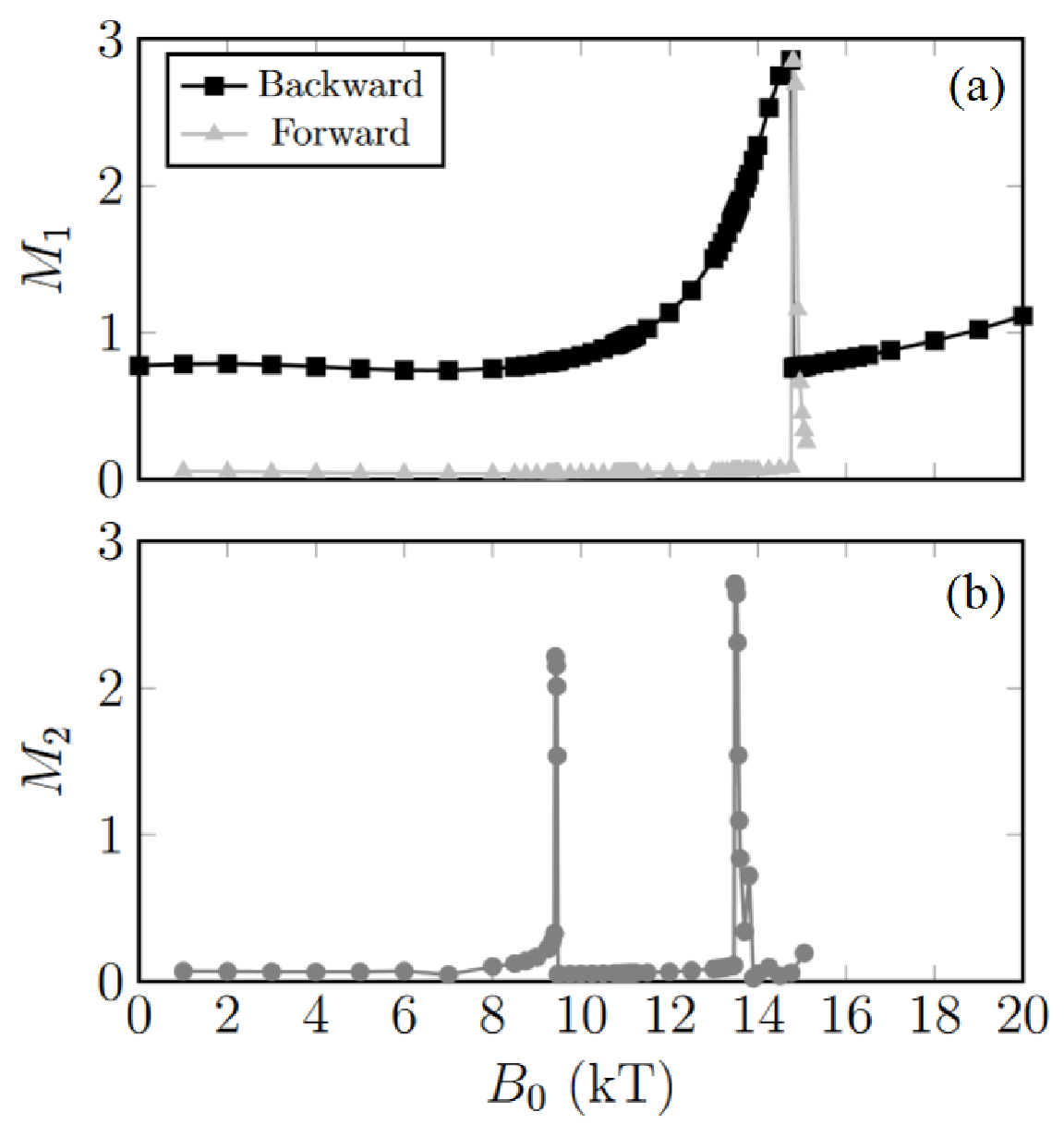}
	\caption{\label{fig:mvals} Normalized growth rates $M=\gamma/\gamma_R$ of dominant scattering channels from warm-fluid theory when waves propagate at $60^\circ$ with respect to $\mathbf{B}_0$. (a) Primary backward (squares) and forward (triangles) scattering peak near \mbox{$B_0=15$ kT} due to mediation by electron-cyclotron-like F waves. At larger $B_0$, F-wave frequency exceeds half the laser frequency, so resonance conditions can only be satisfied for Langmuir-like P waves. (b) Secondary forward scattering (circles) sharply peaks near \mbox{$B_0=9.5$ and 13.5 kT}, which are pumped by the backscattered light and are mediated by the F wave.}
\end{figure}

The choice of simulation parameters and the interpretation of simulation results are guided by fluid theories of coherent magnetized three-wave interactions. \cite{shi2017three,shi2018laser,shi2019three,shi2021theories&simulations} Although kinetic simulations differ from fluid theories, predictions of fluid theories have been found to be indicative of kinetic simulation results. \cite{jia2017kinetic,edwards2019laser}
When pumped by a laser with dimensionless amplitude $a=eE/m_e\omega c$, the undamped growth rate of a resonant three-wave interaction can be written as $\gamma_0=\gamma_R M$, where the dimensionless value $M$ encapsulates magnetization effects and $\gamma_R=\sqrt{\omega\omega_p}a/2$ is the growth rate of unmagnetized Raman backscattering in a cold plasma of the same density. In our setup \mbox{$\gamma_R\approx33$ Trad/s}, where \mbox{Trad} is \mbox{$10^{12}$ rad}, so instabilities typically grow on sub-picosecond time scales.
The $M$ values reported in this paper are from a warm-fluid theory \cite{shi2019three} with polytropic index $\xi=3$. The polytropic index is such that the internal plasma energy is $\epsilon=\xi k_BT$. In magnetized plasmas, where electrons are confined along the magnetic field, $\xi=1$ might be more appropriate, but changing the value of $\xi$ does not qualitatively affect the behaviors of $M$. 
In Figure~\ref{fig:mvals}, the normalized growth rate $M$ is plotted for multiple scattering channels across a range of magnetic field strengths. 
The $M$ value is invariant when plasma conditions and wave 4-momentum are scaled in conjunction. \cite{shi2019amplification,shi2021theories&simulations}
The growth rates of primary instabilities ($M_1$) are shown in Figure~\ref{fig:mvals}(a) and the growth rates of leading secondary instabilities ($M_2$) are shown in Figure~\ref{fig:mvals}(b).
In magnetized plasma, there are usually multiple resonant wave triads, and here we only plot the triads with the largest $M$ values.
The major peaks will be discussed below in the context of simulation results.

\section{\label{sec:plasmawaves}Spontaneous excitation of magnetized plasma waves}
When a laser beam propagates through a plasma, thermal and other fluctuations seed the growth of parametric instabilities. In PIC simulations, statistical noise due to particle sampling mimics the effects of physical fluctuations, and the laser scatters spontaneously. The scattering is dominated by coherent three-wave interactions whenever the resonance conditions $k_1^\mu=k_2^\mu+k_3^\mu$ are satisfied, where $k^\mu=(\omega,\mathbf{k})$ is the wave 4-momentum and $\omega=\omega(\mathbf{k})$ satisfies the wave dispersion relation.
The highest-frequency wave $k_1$ is the pump wave, and waves $k_2$ and $k_3$ are two daughter waves. In our setup, waves $k_1$ and $k_2$ are mostly transverse with a small longitudinal component, whereas wave $k_3$ is mostly longitudinal with small transverse components. 
Since eigenmodes in magnetized plasmas are neither purely transverse nor purely longitudinal when $\mathbf{B}_0$ is at an oblique angle with $\mathbf{k}$, analyzing a single field component is sufficient to reveal all waves that are present in the system. 

\begin{figure*}
	\includegraphics[width=0.85\textwidth]{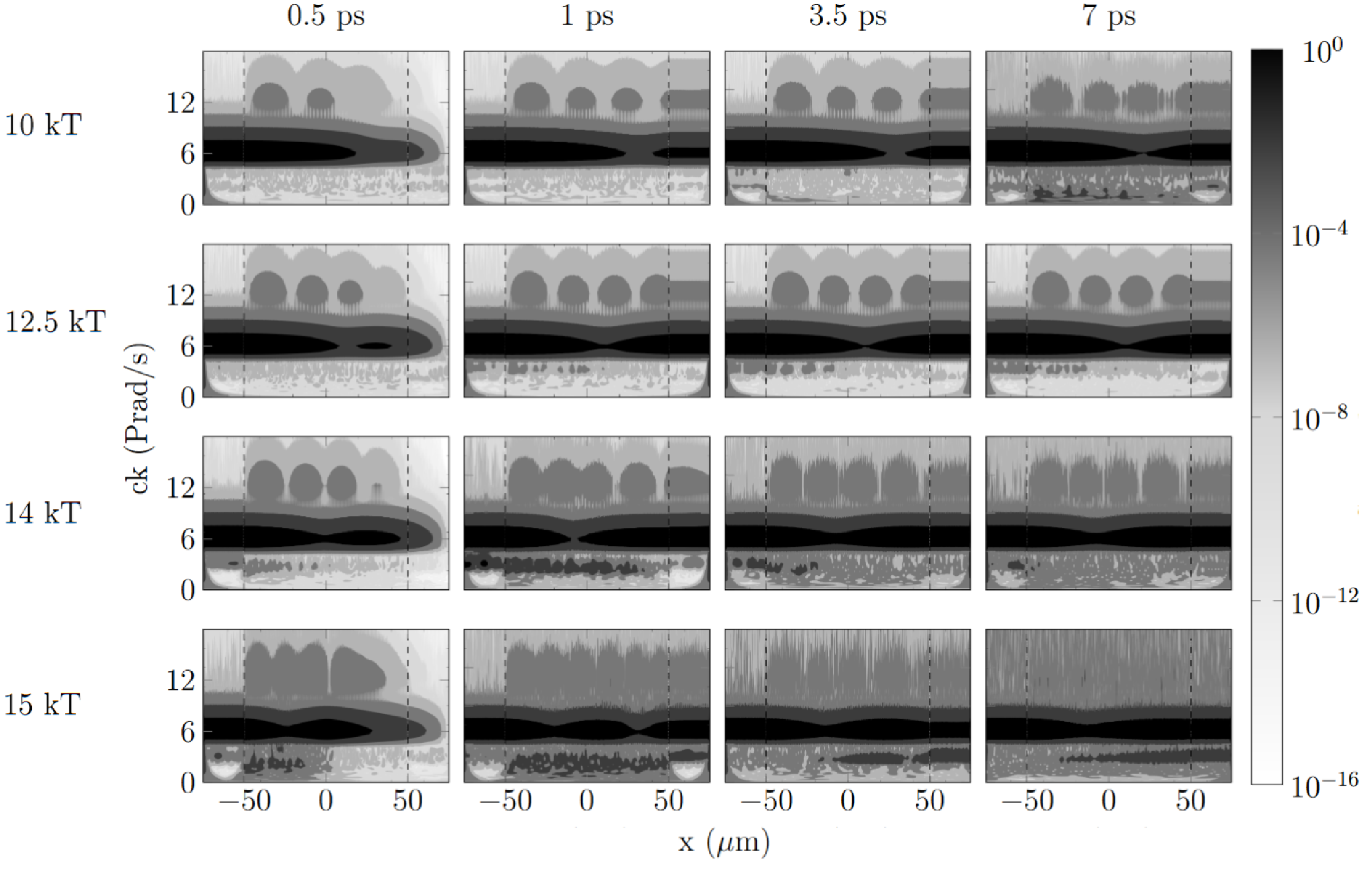}
	\caption{\label{fig:CWTs_all} Noise-subtracted normalized CWT power spectra of $E_y$ for four background magnetic field strengths $B_0$. Each horizontal band corresponds to a coherently excited wave. Common to all cases is backscattering via a Langmuir-like P wave, whereas scattering via electron-cyclotron-like F wave is sensitive to $B_0$.
	In the secondary-forward scenario (SF2, \mbox{$B_0=10$ kT}), backscattered light serves as a pump to grow strong F waves via secondary forward scattering. The process is slow and produces a complex spectrum in the low-$k$ region at \mbox{$t=7$ ps}.
	In the reference scenario (SB0, \mbox{$B_0=12.5$ kT}), which is representative of most $B_0$ values, no F wave is excited to large amplitude. 
	In the primary-backward scenario (SB1, \mbox{$B_0=14$ kT}), backscattering via the F wave becomes a dominant resonance, which leads to a lower-$k$ band.
	In the primary-forward scenario (SF1, \mbox{$B_0=15$ kT}), forward scattering via the F wave is a dominant resonance.}
\end{figure*}

To identify the excited waves, we use continuous wavelet transform (CWT) to analyze snapshots of electromagnetic fields in our simulations. The CWT allows for extraction of characteristic wave vectors, similar to the Fourier transform, while also spatially distinguishing the plasma region from the vacuum regions where the wave vectors are different.  
To reduce effects of PIC noise in CWT, we run each simulation twice: once with the laser and once without. Subtracting the laser-off CWT power spectrum from the laser-on power spectrum removes laser-independent waves, which are excited due to PIC noise before the laser arrives. 
Then, in the noise-subtracted CWT $|S(x,k)|^2$, each horizontal band can be associated with a coherently excited wave. The CWTs for $E_y$ are shown in Figure~\ref{fig:CWTs_all}; those of other field components are similar. 
The dashed vertical lines mark the initial plasma boundaries, and the U shape near the bottom corners is an artifact of CWTs at small wave vectors.

Common to all scenarios shown in Figure \ref{fig:CWTs_all} are three wave bands corresponding to the incident laser at \mbox{$ck_1\approx5.3$ Prad/s}, the magnetized Langmuir-like P wave at \mbox{$ck_3\approx10.2$ Prad/s}, and the resultant backscattered light at \mbox{$ck_2\approx4.9$ Prad/s}, where \mbox{Prad} is \mbox{$10^{15}$ rad}.
These daughter waves are the result of magnetized Raman-like scattering, \cite{grebogi1980brillouin,shi2017laser,jia2017kinetic} whose resonance conditions can always be satisfied.
At given plasma density and temperature, changing $B_0$ only slightly affects the P wave as long as $\omega_{c}$ is not close to $\omega_p$. Consequently, the kinematics of the P-wave mediated scattering only weakly depends on $B_0$, and the growth rate $M\sim O(1)$ varies slightly with the magnetic field.

\begin{figure*}
	\includegraphics[width=0.8\textwidth]{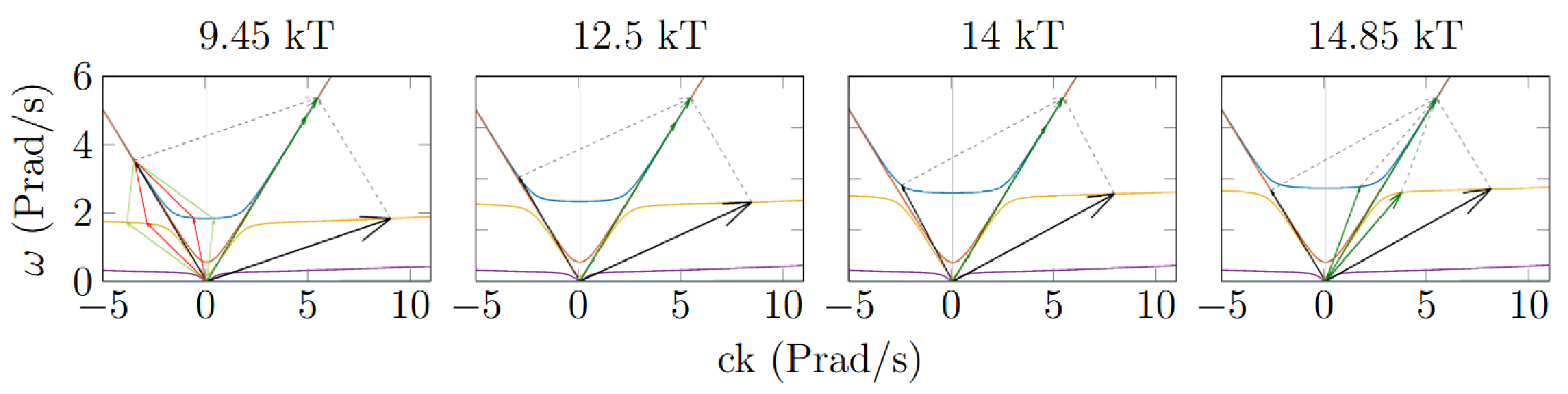}
	\caption{\label{fig:dispersionrelations} Resonant wave triads for F-wave mediated scattering from warm-fluid theory when waves propagate at $60^\circ$ with respect to $\mathbf{B}_0$. The dispersion relation shows four of the eigenmodes relevant for our simulations. From top to bottom, they are the right-handed elliptically polarized light (blue), the left-handed elliptically polarized light (orange), the electron-cyclotron-like F wave (yellow), and the Langmuir-like P wave (purple). 
	Resonant wave triads are identified by parallelograms of energy-momentum conservation. Primary forward scattering is shown by green arrows, primary back scattering is shown by black arrows, and secondary forward scattering is shown in red.}
\end{figure*}

In addition, coherent scattering can be mediated by the electron-cyclotron-like F wave, which is present only when $B_0$ is nonzero and the propagation angle is oblique. For the four representative scenarios shown in Figure~\ref{fig:CWTs_all}, resonant wave triads that involve F waves are identified by parallelograms in Figure~\ref{fig:dispersionrelations}. The arrow along the diagonal of the parallelogram represents the pump wave, while the edges represent the daughter waves. In the case of primary scattering, the pump is the incident laser. In the case of secondary scattering, a daughter wave from the primary scattering acts as the pump. 

\begin{figure}[b]
	\includegraphics[width=0.45\textwidth]{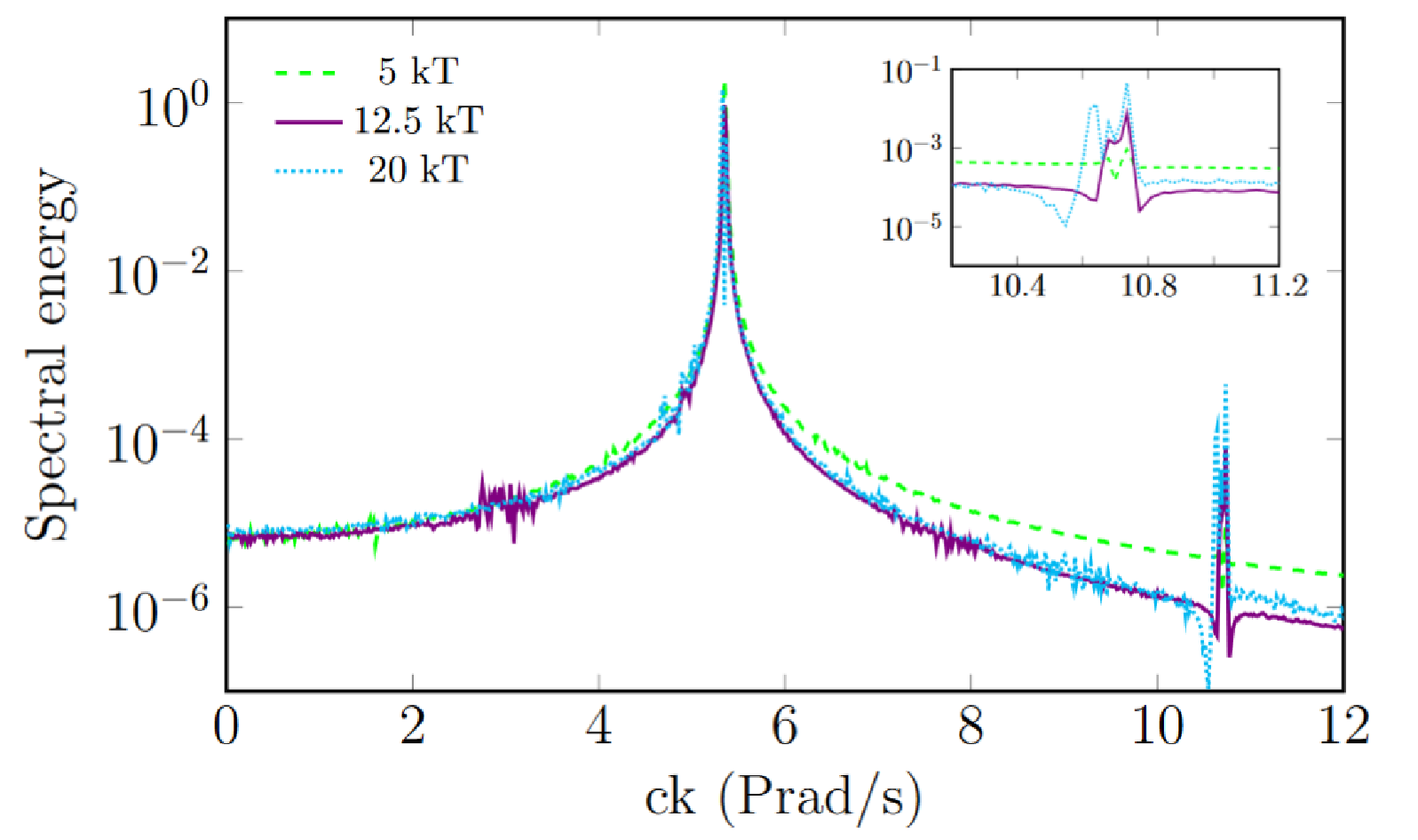}
	\caption{\label{fig:FFTs} Normalized Fourier power spectra of $E_y$ in the plasma region for cases without strong F wave coupling. The main peak is due to the incident and backscattered light from the Langmuir-like P wave. The peak near \mbox{$ck=10$ Prad/s} is due to the P wave, which has a weak dependence on $B_0$. At \mbox{$B_0=12.5$ kT}, the features near \mbox{$ck_2=3$ Prad/s} and \mbox{$ck_3=8$ Prad/s} are due to backscattering mediated by electron-cyclotron-like F wave, which is substantially weaker. }
\end{figure}

In the SB0 scenario, backscattering via the F wave is a subdominant resonance (Figure \ref{fig:CWTs_all}, row 2). This scenario is the simplest and most common, and is illustrated using \mbox{$B_0=12.5$ kT} as an example. 
Compared to backscattering mediated by the P wave, F-wave mediated backscattering usually has a smaller growth rate and hence plays a subdominant role. In the particular example of \mbox{$B_0=12.5$ kT}, the P-wave mediated backscattering has an undamped growth rate of $M\approx0.7$, while the F-wave mediated backscattering has $M\approx1.2$. However, as will be shown in Sec.~\ref{sec:vdfs}, the F wave experiences larger collisionless damping and therefore a smaller overall growth. 
In simulations, to illustrate that P-wave mediated scattering is indeed dominant at most $B_0$ values, we overlay the Fourier power spectra of $E_y$ in the plasma region for three very different values of $B_0$ in Figure~\ref{fig:FFTs}. Apart from the peak due to the incident laser at \mbox{$ck_1\approx5.3$ Prad/s}, the most prominent peak is at \mbox{$ck_3\approx10.5$ Prad/s}, which gives rise to backscattered $ck_2$ that largely overlaps with $ck_1$. The wave vector of this resonance has little dependence on $B_0$ and is close to the expected P wave. 
Notice that the subdominant F wave is also visible in the Fourier spectra. At \mbox{$B_0=12.5$ kT}, the feature near \mbox{$ck_2\approx3$ Prad/s} corresponds to the expected backscattering mediated by the F wave, whose resonant wave vector gives rise to the feature near \mbox{$ck_3\approx8$ Prad/s}. 
At larger $B_0$, the cyclotron frequency increases, so the resonant $ck_2$ and $ck_3$ are smaller.

Second, in the SB1 scenario, backscattering via the F wave becomes a dominant resonance (Figure \ref{fig:CWTs_all}, row 3). This occurs near \mbox{$B_0=14$ kT} in our setup, where backscattering mediated by the F wave has a normalized growth rate of $M\approx2.3$. In comparison, the backscattering mediated by the P wave has $M\approx0.7$, which is much weaker.
To understand the origin of the strong F-wave mediated resonance, note that in Figure~\ref{fig:dispersionrelations} the group velocity of the backscattered wave is noticeably lower than other cases. A lower group velocity indicates that a larger fraction of the wave energy is carried in the form of kinetic energy rather than field energy. Consequently, the wave interacts more strongly with the plasma and the three-wave coupling is larger. 
Signatures of F-wave mediated strong backscattering can be seen near \mbox{$t=1$ ps} in the CWT, where an additional low-$k$ band at \mbox{$ck\approx2.4$ Prad/s} becomes pronounced. 

In the third scenario (SF1), F-wave mediated forward scattering is a dominant resonance (Figure~\ref{fig:CWTs_all}, row 4), which we observe near \mbox{$B_0=15$ kT}. In this scenario, in addition to being backscattered via the P wave with a normalized growth rate $M\approx 0.8$, the laser is also strongly forward scattered via the F wave with $M\approx 2.7$, which is unusually large for forward-scattering.
To understand the strong forward scattering, which never occurs in the unmagnetized case, notice that the wave dispersion relation near $\omega_c$ allows for a forward scattering parallelogram with a substantial area. The large area correlates with a large phase-space volume of the scattering process, which supports a large scattering cross section. More importantly, the coupling is strong because both daughter waves have slow group velocities (Figure~\ref{fig:dispersionrelations}) and thus a substantial fraction of energy carried in the form of particle motion.
In this scenario, a total of three nearly electrostatic plasma waves are excited to large amplitudes. Two of them are daughter waves of F-wave mediated forward scattering, which appear in the CWT as two low-$k$ bands.

Finally, in the SF2 scenario, the backscattered laser serves as a pump to induce strong secondary forward scattering via the F wave (Figure \ref{fig:CWTs_all}, row 1). In our setup, this occurs near \mbox{$B_0=9.5$ kT} and \mbox{13.5 kT}.
In the former case, the pump of the secondary forward scattering, whose undamped growth rate is $M\approx2.2$, is the daughter wave of F-wave mediated backscattering. 
In comparison, at \mbox{$B_0=13.5$ kT}, the pump of the secondary forward scattering, whose undamped growth rate is $M\approx2.7$, is the daughter wave of P-wave mediated backscattering. 
In both cases, the secondary pumps have frequencies close to twice the F-wave frequency, and they undergo strong F-wave mediated forward scattering for similar reasons as in the SF1 scenario. We focus on the \mbox{$B_0=10$ kT} case to illustrate the SF2 scenario, and the case when \mbox{$B_0=13.5$ kT} is similar. 
Since the pump is now the backscattered light, which needs to build up its amplitude from the primary process, the secondary forward resonance is weaker and takes longer time to develop, as can be seen from Figure~\ref{fig:CWTs_all}. Unlike the scenarios at \mbox{$B_0=14$ kT} and \mbox{15 kT}, which produce significant additional scattering within \mbox{$t=1$ ps}, the \mbox{10 kT} scenario does not show a similar increase until after \mbox{$t=3.5$ ps}. By \mbox{$t=7$ ps}, the plasma region is filled with many waves spanning multiple wave vectors, suggesting a more complicated series of wave interactions than in previous scenarios.
Tertiary and higher order scattering processes are possible, but they do not significantly affect the results discussed here.

\section{\label{sec:vdfs}Broadening of velocity distribution functions due to collisionless wave damping}
Under most background magnetic field strengths, we encounter the SB0 scenario where P-wave mediated interaction is the only strong resonance. In this scenario, no significant broadening of $f(\mathbf{v})$ is observed. This is expected because the quiver velocity of electrons in the laser is $v_q/c\simeq eE/m_e\omega_1 c\approx 0.04$, the phase velocity of the P wave is $v_p/c\simeq\omega_p/ck_3\approx0.05$, and the trapping velocity of electrons in an electrostatic wave of the same field strength is $v_t\simeq\sqrt{eE/m_e k}\approx 0.04c$, which are all comparable to the electron thermal velocity $v_T=\sqrt{2k_B T_e/m_e}\approx0.06c$. 
As a result, the spread of $f(\mathbf{v})$ at later time remains similar to its initial spread. 
Compared to the unmagnetized case, where collisionless damping of a single electrostatic wave leads to a plateau in $f(v_x)$ near $v_x=v_p$, the $z$ component of $\mathbf{B}_0$ steers trapped particles away from wave-particle resonance and leads to a more symmetric broadening of the distribution function in the $x$-$y$ plane.

\begin{figure}
	\includegraphics[width=0.45\textwidth]{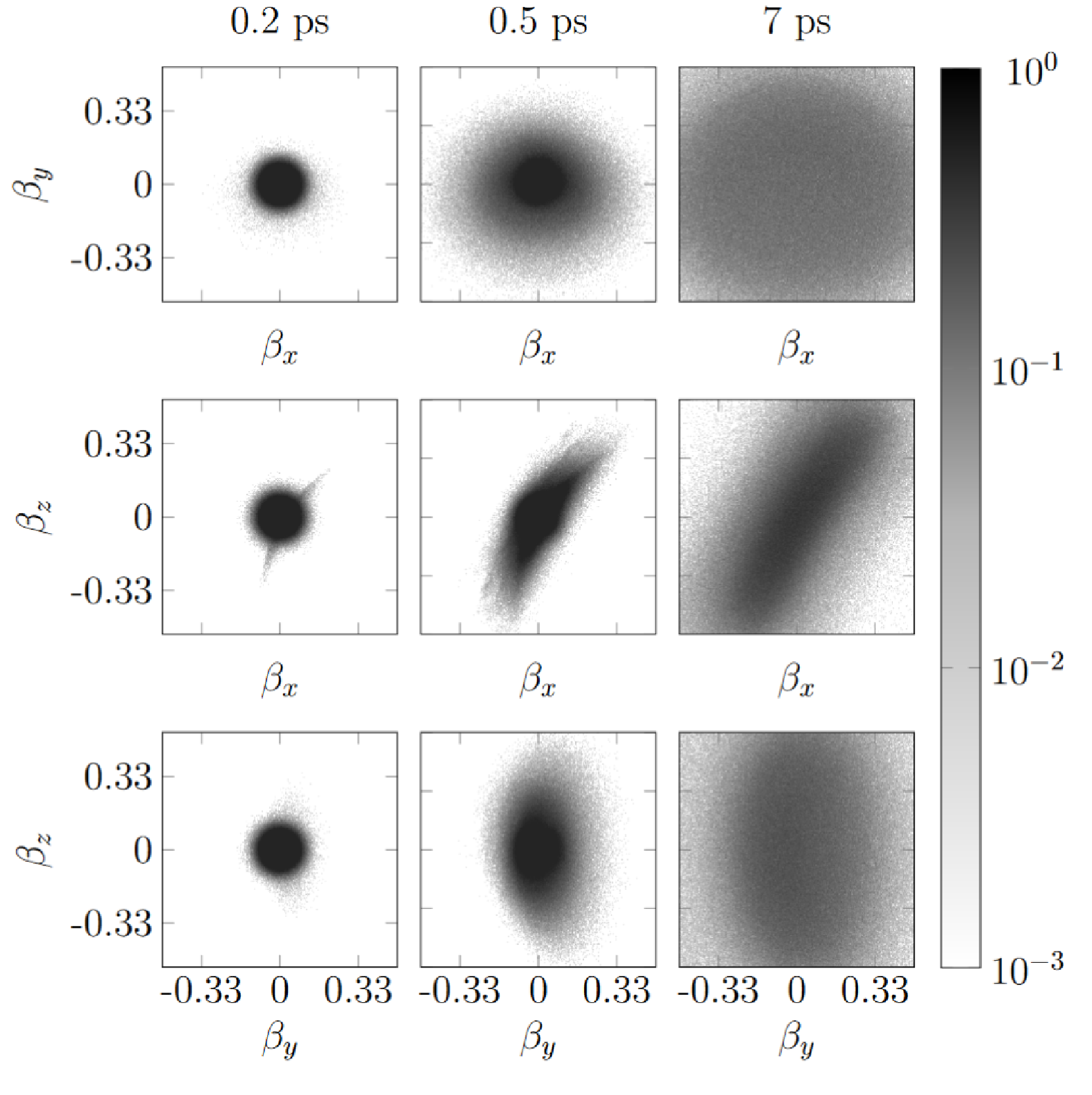}
	\caption{\label{fig:2dVDFs_15kT} Normalized spatially integrated 2D electron velocity distribution functions ($\beta=v/c$) for the primary-forward scenario (SF1, \mbox{$B_0=15$ kT}). Immediately after the laser enters the plasma (\mbox{$t=0.2$ ps}), a feature develops in $f(v_x,v_z)$ along $\mathbf{B}_0$ and curves towards the wave propagation direction. Before the laser front exits the plasma (\mbox{$t=0.5$ ps}), the feature already expands substantially and anisotropic heating becomes apparent in $f(v_x,v_z)$ and $f(v_y,v_z)$. At later time (\mbox{$t=7$ ps}), the distribution function is significantly broadened and approaches an ellipsoid whose major axis is along $\mathbf{B}_0$.}
\end{figure}

\begin{figure}[t]
	\includegraphics[width=0.45\textwidth]{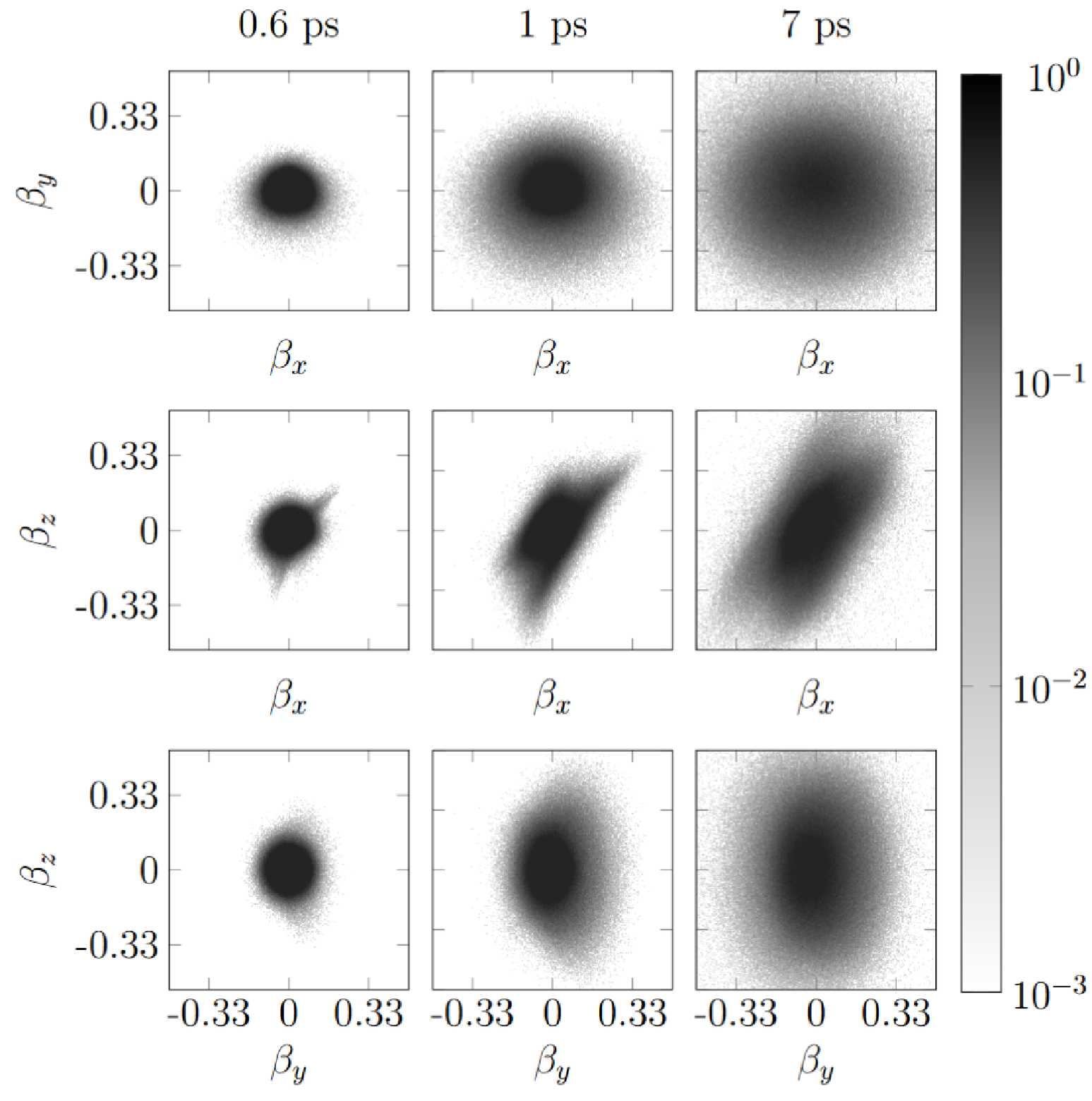}
	\caption{\label{fig:2dVDF_14kT}Normalized 2D electron velocity distributions for the primary-backward scenario (SB1, \mbox{$B_0=14$ kT}). By \mbox{$t=0.6$ ps}, the curved feature due to excitation of the electron-cyclotron-like F wave becomes visible in $f(v_x,v_z)$. Shortly afterwards (\mbox{$t=1$ ps}), another feature whose centroid is shifted towards a smaller $v_x$ also becomes visible. This additional feature is likely due to excitation of the Langmuir-like P wave, which also has large growth rate and propagates in the $+x$ direction. At later time, the two features merge and the final distribution approaches an ellipsoid along $\mathbf{B}_0$.}
\end{figure}

\begin{figure}[t]
	\includegraphics[width=0.45\textwidth]{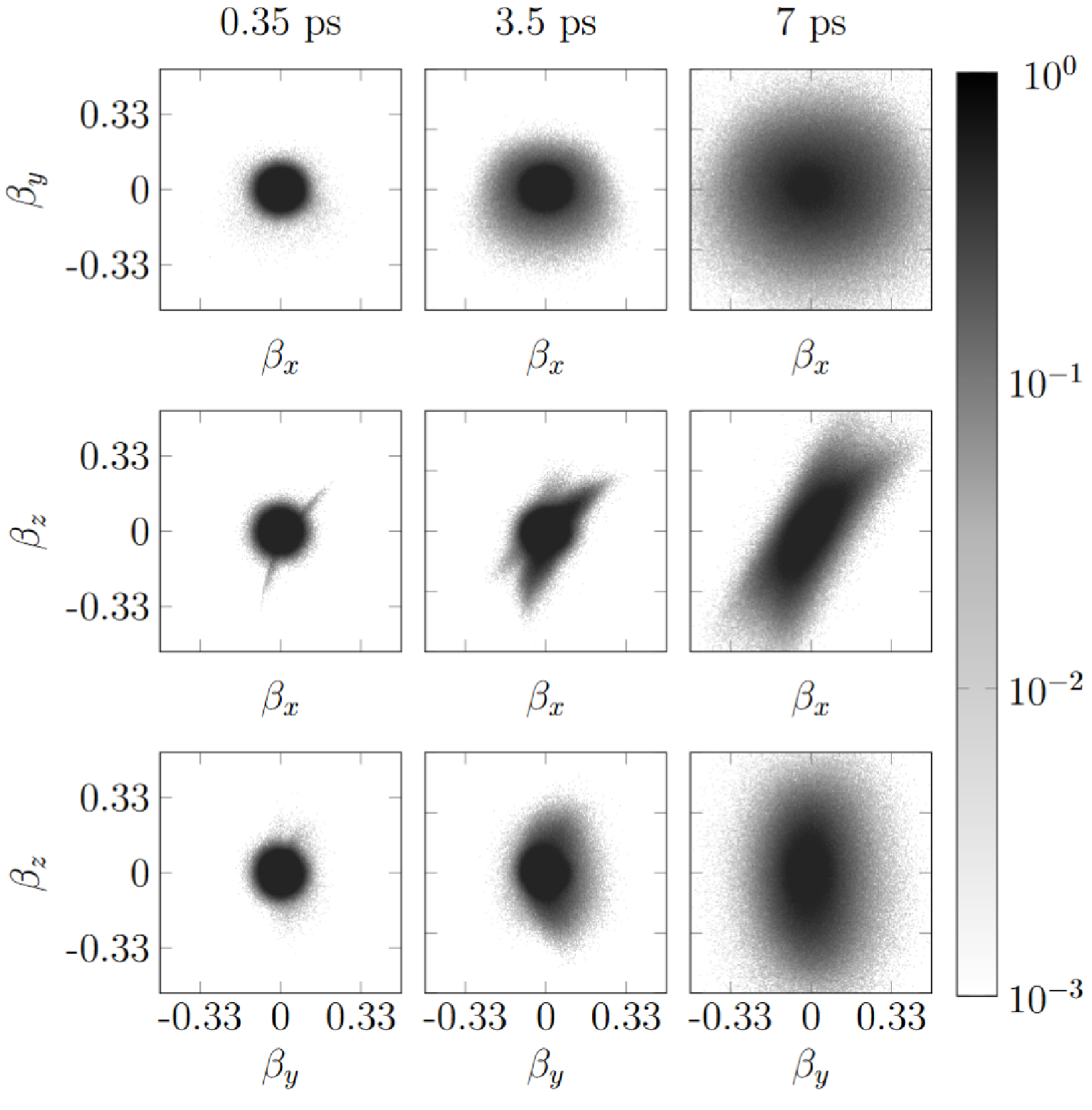}
	\caption{\label{fig:2dVDFs_10kT}Normalized 2D electron velocity distributions for the secondary-forward scenario (SF2, \mbox{$B_0=10$ kT}). At early time (\mbox{$t=0.35$ ps}), a curved feature develops in $f(v_x,v_z)$ due to excitation of an F wave propagating in the $+x$ direction. After backscattered light reaches large amplitude, it pumps another F wave in the $-x$ direction, leading to a fainter feature with opposite convexity by \mbox{$t=3.5$ ps}. At \mbox{$t=7$ ps}, the distribution resembles that shown in Figure~\ref{fig:2dVDF_14kT}, but the broadening has not saturated in this case.}
\end{figure}

Much more dramatic broadening of the distribution function occurs when large amplitude electron-cyclotron-like F waves are also excited (Figures~\ref{fig:2dVDFs_15kT}-\ref{fig:2dVDFs_10kT}), and the broadening of the distribution function is correlated with the development of the F waves.
Three mechanisms likely contribute to this phenomenon. 
First, in the presence of multiple resonant wave triads, the dynamics are in general chaotic,~\cite{kueny1995nonlinear} leading to strong diffusion in the electron phase space. 
Second, electron orbits due to a single F wave are already chaotic due to competing wave motion and gyromotion.~\cite{dodin2011surfatron} Since the frequency of F wave is close to $\omega_c$, the dynamics are more complicated than the stochastic heating mechanism\cite{Karney78,Karney79} where gyromotion dominates wave motion, as well as the surfatron acceleration mechanism\cite{Sagdeev73,Dawson83} where wave motion dominates gyromotion.  
Third, oblique F waves have non-negligible transverse components and a subluminal phase velocity $v_p<c$. Consequently, when boosted to the reference frame traveling at the wave phase velocity, particles experience a secular acceleration due to the transverse wave electric field, which has a component parallel to the direction of the magnetic field. The secular electric field accelerates particles to velocities that are much larger than the wave quiver velocity. 
The last mechanism probably contributes the most to anisotropic distribution functions we observe in simulations, which are significantly elongated along the direction of the magnetic field. However, the exact contribution of each mechanism remains to be further investigated in the future.

In the SF1 scenario (\mbox{$B_0=15$ kT}), where the interaction is dominated by F wave mediated forward scattering, a single feature is developed in velocity space (Figure \ref{fig:2dVDFs_15kT}).
Since the instability has a large growth rate, the F wave develops a substantial amplitude by \mbox{$t=0.2$ ps}, and a diagonal feature of high-velocity electrons emerges in $f(v_x,v_z)$. This feature primarily expands along the direction of $\mathbf{B}_0$, which is in the $x$-$z$ plane and at $60^\circ$ with respect to the $x$ axis. Additionally, the feature curves towards the $+v_x$ direction, which is correlated with the forward wave propagation direction. 
As time progresses, the distribution function continues to elongate along $\mathbf{B}_0$ and broadens across the magnetic field. At the same time, the convexity becomes less pronounced, and the distribution function approaches an ellipsoid with its major axis along $\mathbf{B}_0$. The final $f(\mathbf{v})$ is significantly broader than the initial distribution function.

In the SB1 scenario (\mbox{$B_0=14$ kT}), where F-wave mediated backscattering dominates, a double feature develops in the velocity space (Figure \ref{fig:2dVDF_14kT}).
At early time (\mbox{$t=0.6$ ps}), a curved feature develops in accordance with the growth of the forward-propagating F wave. This is similar to what happens in SF1, except that the feature takes longer time to emerge because the growth rate is smaller. 
Shortly afterwards (\mbox{$t=1$ ps}), another feature appears in $f(v_x,v_z)$. This feature also expands along the magnetic field, but its centroid is shifted to smaller $v_x$. This additional feature is not pronounced in the SF1 scenario and is likely due to the excitation of the P wave, whose growth rate is also appreciable. Notice that the resonant P wave has a smaller phase velocity, which is correlated with the shift towards smaller $v_x$.
After \mbox{1 ps}, the two features merge into an ellipsoid whose major axis is again along $\mathbf{B}_0$. The final distribution in this case is less broad than in the SF1 scenario but still much hotter than the initial distribution.

Finally, in the SF2 scenario, an opposite double feature is developed in the velocity space (Figure \ref{fig:2dVDFs_10kT}). 
At early time (\mbox{$t=0.35$ ps}) when \mbox{$B_0=10$ kT}, the incident laser pumps an F wave that propagates in the $+x$ direction. The growth of this F wave leads to an $f(v_x,v_z)$ that is elongated along the magnetic field and curved towards the $+v_x$ direction.
After the backscattered light grows to large amplitude, it starts to pump secondary instabilities and excites another F wave that propagates in the $-x$ direction. Since this is a secondary process, it takes longer time to develop. By \mbox{$t=3.5$ ps}, another feature that extends along $\mathbf{B}_0$ becomes clearly visible in $f(v_x,v_z)$. Notice that this fainter feature extends towards the $-v_x$ direction, which is correlated with the opposite wave propagation direction. 
At later time, the two features again merge and the final distribution function approaches an ellipsoid. The extent of  $f(\mathbf{v})$ by \mbox{$t=7$ ps} is comparable to that in the SB1 scenario. However, in this case, the expansion of the distribution function has not saturated by the end of our simulation. 
What happens for the SF2 scenario near \mbox{$B_0=13.5$ kT} is similar.

\begin{figure}[t]
	\includegraphics[width=0.43\textwidth]{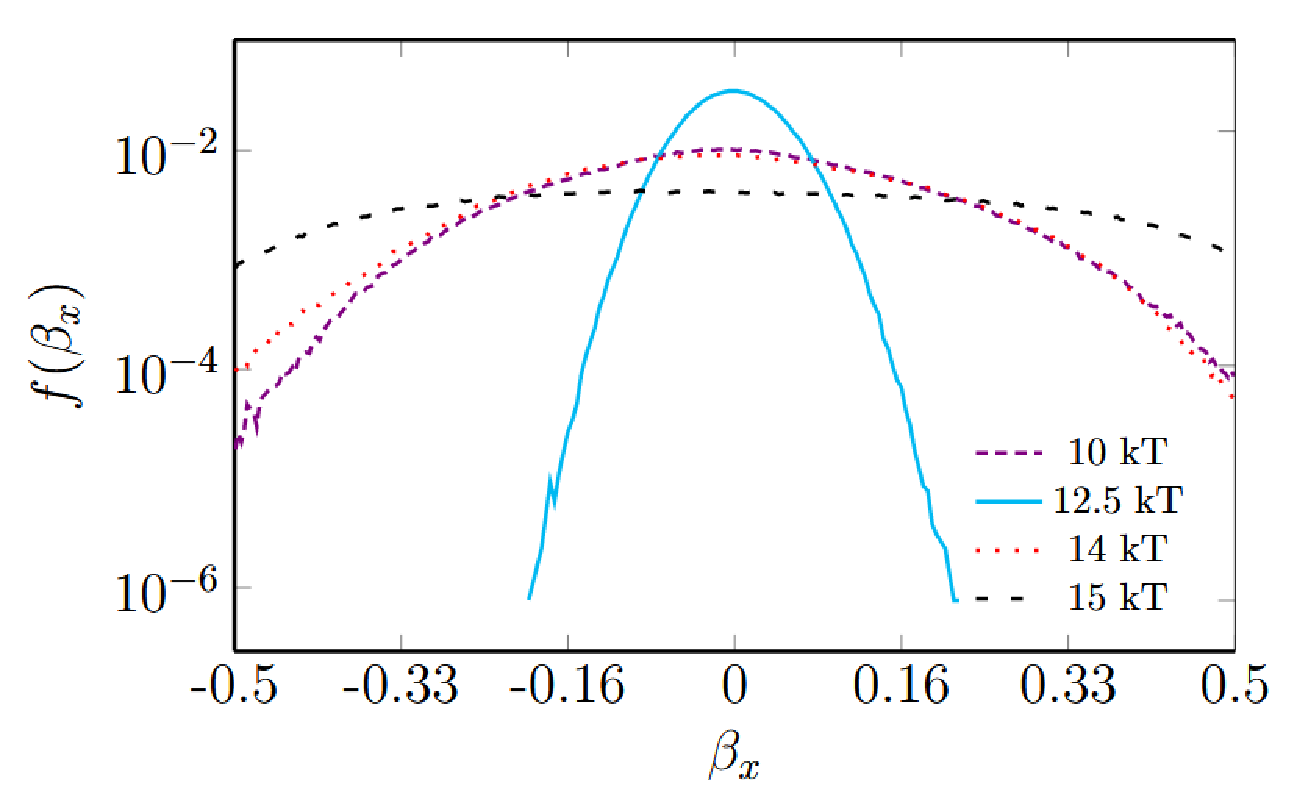}
	\caption{\label{fig:phasevel} Normalized electron velocity distribution functions \mbox{$f(\beta_x)$} for all four scenarios at \mbox{$t=7$ ps}. In the SB0 scenario (\mbox{$B_0=12.5$ kT}, solid blue), the final distribution function is close to the initial. For the SF1 scenario (\mbox{$B_0=15$ kT}, large dashed black), the bulk of $f(v_x)$ is strongly heated and non-Maxwellian tails are also prominent. The distributions in scenarios SB1 (\mbox{$B_0=14$ kT}, dotted red) and SF2 (\mbox{$B_0=10$ kT}, small dashed purple) are similar, but the broadening has not saturated in the later scenario.}
\end{figure}

To facilitate comparison across different $B_0$ values, we overlay one dimensional distribution functions $f(v_x)$ at \mbox{$t=7$ ps} in Figure~\ref{fig:phasevel}. 
In the SB0 scenario, the final $f(v_x)$ is close to the initial distribution function. In contrast, in scenarios where large-amplitude F waves are excited, the distribution function is significantly broadened and the high-density hot-temperature tails are pronounced.
The broadening of $f(v_x)$ is most significant in the SF1 scenario (\mbox{$B_0=15$ kT}), where the forward F wave has an exceptionally large growth rate and a high phase velocity. 
The large-amplitude F waves that mediate strong backscattering, as seen in the SB1 scenario near \mbox{$B_0=14$ kT}, also lead to significant heating of the plasma. 
Finally, although broadening is slower in the SF2 scenario, because the amplitude of F waves are smaller compared to the other two scenarios at any given time, the final distribution function also reaches high temperature. 
It is worth emphasizing that the strong heating is purely due to collisionless wave-particle interactions, which are much more effective than the usual collisional heating via inverse Bremsstrahlung.

\section{\label{sec:energy}Enhanced laser energy absorption from wave-particle interactions}
\begin{figure}[b]
	\includegraphics[width=0.38\textwidth]{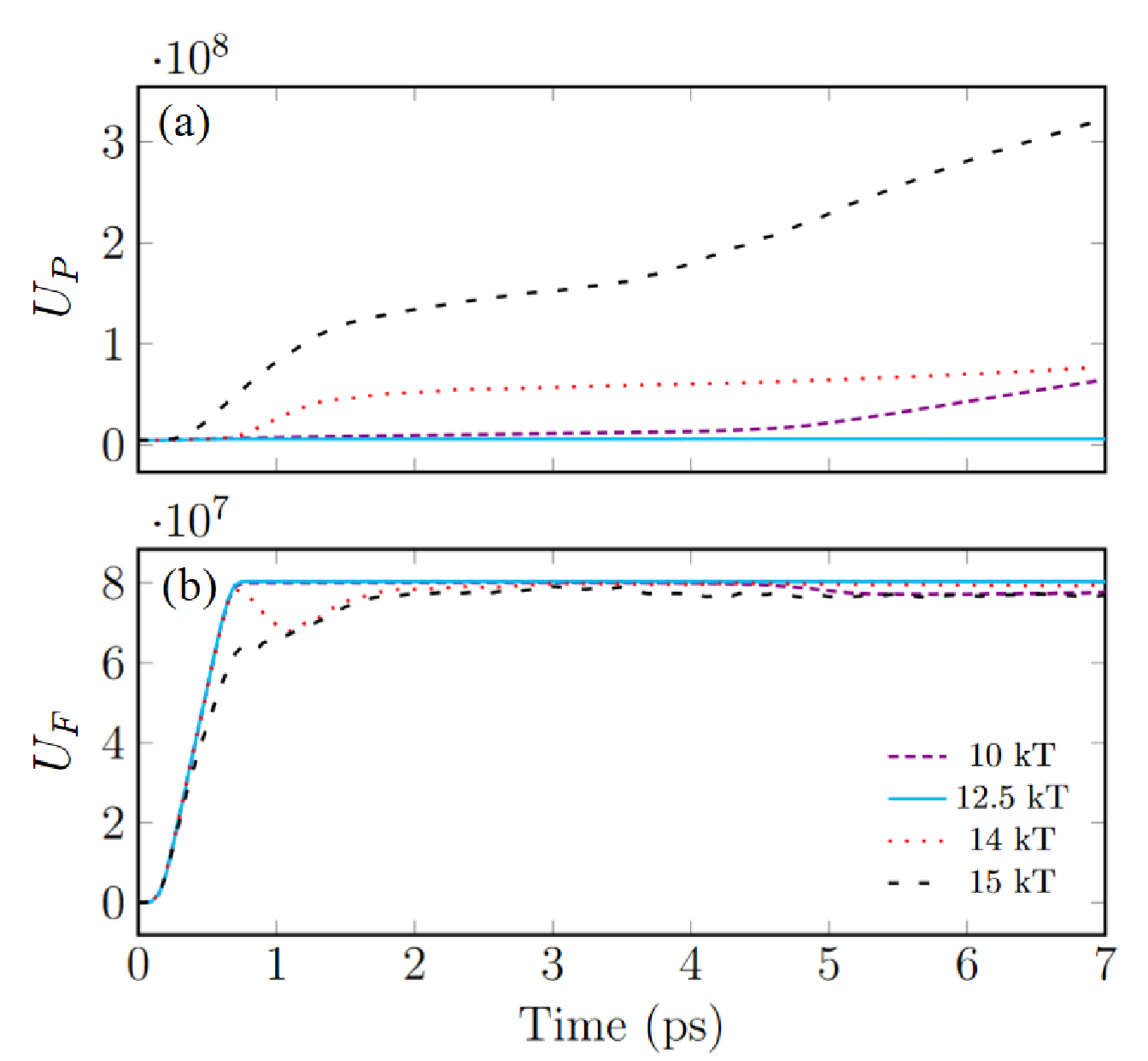}
	\caption{\label{fig:plasen} Areal energy density of particles (a) and fields (b) in units of $\text{J/m}^2$. 
	In the SB0 scenario (\mbox{$B_0=12.5$ kT}, solid blue), which is representative of what happens at most $B_0$ values, plasma energy ($U_P$) and field energy ($U_F$) plateau after the laser fills up the simulation box. 
	In the SB1 scenario (\mbox{$B_0=14$ kT}, dotted red), resonant waves quickly develop and transfer energy from fields to particles, leading to a dip in $U_F$. After the mechanism saturates, the energy plateaus again. 
	In the SF1 scenario (\mbox{$B_0=15$ kT}, large dashed black), strong cyclotron-like resonances are excited, leading to dramatic increase of $U_P$ and substantial absorption of the laser. 
	In the SF2 scenario (\mbox{$B_0=10$ kT}, small dashed purple), after an initial build up of primary waves, $U_P$ starts to increase significantly, which is accompanied by a decrease of $U_F$.
	The SF mechanisms have not saturated by \mbox{7 ps}.}
\end{figure}

\begin{figure}
	\includegraphics[width=0.35\textwidth]{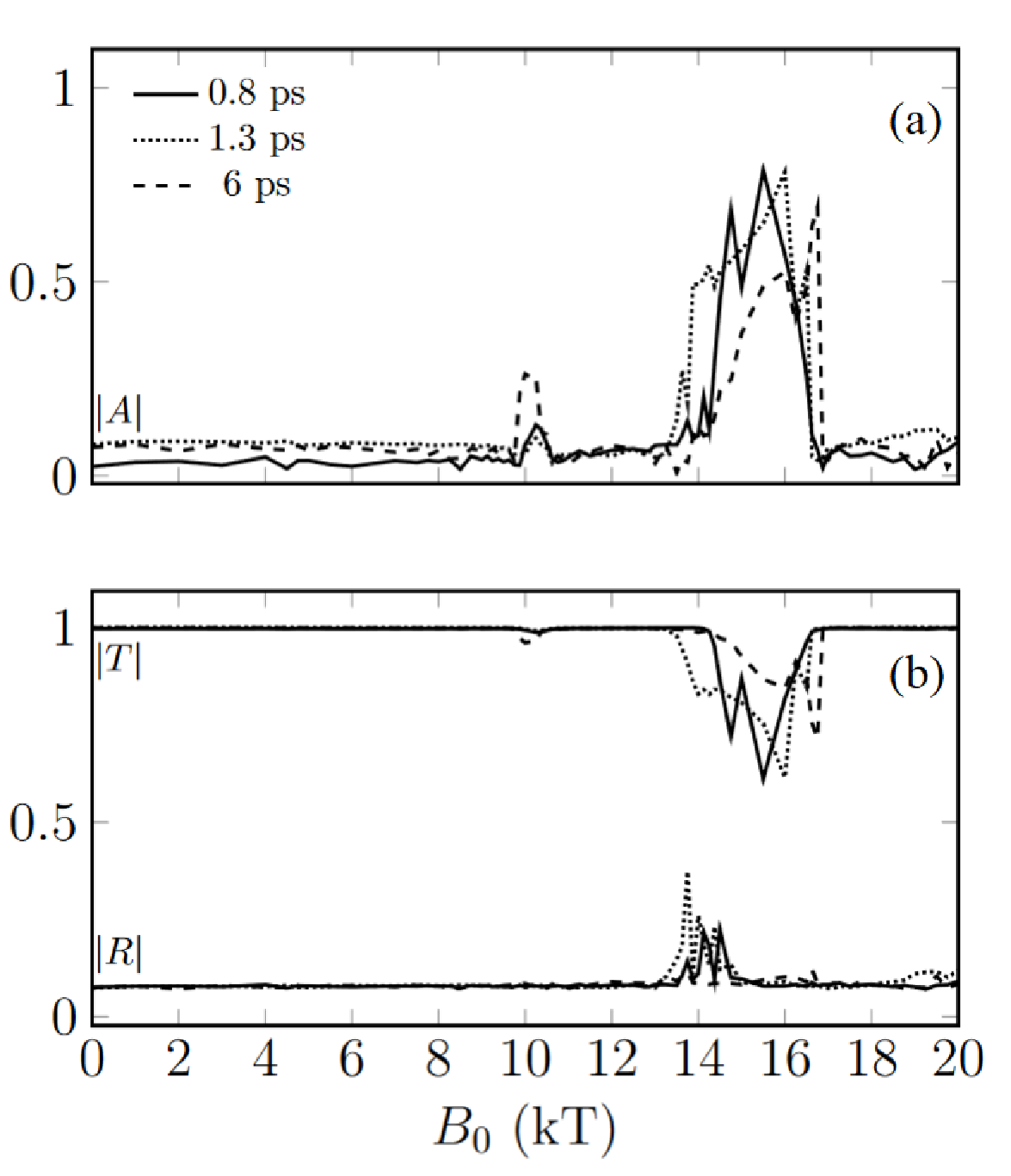}
	\caption{\label{fig:absorption} Absorption ($A$), reflection ($R$), and transmission ($T$) coefficients at three time slices. Absorption peaks at $B_0$ values where electron-cyclotron-like F waves are strongly excited. 
	Near \mbox{$B_0=10$ and 13.5 kT}, absorption is due to secondary forward scattering, which has little effect on $R$. Since this is a secondary process, it takes more time to develop and $A$ increases in time. 
	Near \mbox{$B_0=14$ kT}, absorption is due to primary backscattering, which leads to an increase of $R$. In this scenario, $T$ drops due to both enhanced absorption and reflection. The mechanism saturates later in time.
	Near \mbox{$B_0=15$ kT}, absorption is due to primary forward scattering, so $R$ is unaffected while $T$ substantially decreases. Collisionless absorption is strongest in this scenario and is much larger than what is possible with collisional inverse Bremsstrahlung.}
\end{figure}

To confirm that laser energy is absorbed via collisionless damping of plasma waves, we analyze the energy balance of our simulations in detail. 
The integral form of the energy-conservation law is $\partial_t \int_0^L dx U = S_x(0)-S_x(L)$, where $L$ is the length of our one-dimensional simulation box. 
On the left-hand side of the conservation law, $U=U_P+U_F$ is the total energy density, where $U_P=\sum_s\frac{1}{2}m_s\int d\mathbf{v}f_s(\mathbf{v})\mathbf{v}^2$ is the plasma energy density summed over species and $U_F=\frac{1}{2}\epsilon_0\mathbf{E}^2+\frac{1}{2\mu_0}\mathbf{B}^2$ is the field energy density. The areal energy densities are shown in Figure~\ref{fig:plasen}, for which $U$ is integrated along $x$.
On the right-hand side of the conservation law, the energy flux is $\mathbf{S}=\mathbf{S}_P+\mathbf{S}_F$, where the flux due to particles is $\mathbf{S}_P=\sum_s\frac{1}{2}m_s\int d\mathbf{v}f_s(\mathbf{v})\mathbf{v}^2\mathbf{v}$. Since we include two vacuum gaps in our setup and few particles have left the box by the end of our simulations, $\mathbf{S}_P$ equals to zero on both boundaries. 
Therefore, the energy fluxes at $x=0$ and $x=L$ are entirely due to the field contribution $\mathbf{S}_F=\frac{1}{\mu_0}\mathbf{E}\times\mathbf{B}$. The simulations are one dimensional in space, so the flux is only in the $x$ direction. Moreover, it is easy to separate the left- and right-moving components using Faraday's Law. 
Denoting $E_y^\pm = (E_y\pm cB_z)/2$ and $E_z^\pm = (E_z\mp cB_y)/2$, we can calculate energy flux $S^\pm = \epsilon_0 c [(E_y^\pm)^2 + (E_z^\pm)^2]$ in the vacuum gap, where $\pm$ represent fluxes in positive and negative $x$ directions, respectively. From the energy flux, we calculate reflection ($R$), transmission ($T$), and absorption ($A$) coefficients by
\begin{eqnarray}
    |R|^2 &=& \frac{S^-(0)}{S^+(0)},\\
    |T|^2 &=& \frac{S^+(L)}{S^+(0)},\\
    |A|^2 &=& 1-|T|^2-|R|^2,
\end{eqnarray}
which are shown in Figure \ref{fig:absorption}.
The flux $S^-(L)$ is zero, because the laser enters from the left boundary and no wave is left-propagating at the right boundary. 
Notice that we calculate these coefficients using fluxes at equal time. A more rigorous calculation could uses fluxes with proper time retardation. However, since we are dealing with multiple waves with different group velocities, the retarded-time definition is unnecessarily complicated and does not qualitatively change the physical picture.

As the laser beam enters the simulation box, the influx leads to an increase of energy inside the box. During this process, $U_F$ grows while $U_P$ only slightly increases, because the electron quiver velocity in the laser field is smaller than the thermal velocity.
After the laser front exits, the box is filled up. If no additional process occurs, then the outgoing flux balances the incoming influx, and the total energy plateaus in ideally collisionless simulations. 
In our simulations, field energy indeed remains roughly constant in the absence of strong F-wave resonances, but the plasma energy decreases slowly over time due to spurious radiative cooling from numerical noise. Therefore, any rise in plasma energy after the initial increase is attributed solely to resonant wave-particle interactions.

The simple energy balance picture is changed when waves are excited resonantly. 
In the most common SB0 scenario, where only a Langmuir-like P wave is excited, backscattering via the P wave adds to the reflection from the vacuum-plasma boundary, so a small fraction of energy leaves the simulation box from the left. Once excited, the P wave undergoes collisionless damping, which leads to a slight broadening of the distribution function and increase of the plasma energy.

More interesting effects occur when the electron-cyclotron-like F waves are also excited. 
In the SB1 scenario, a strong F-wave mediated resonance scatters laser in the backward direction. The enhanced backscattering is correlated with an increased reflectivity near \mbox{$B_0=14$ kT}, as shown in Figure~\ref{fig:absorption}. After the F wave grows to sufficient amplitudes, it is consumed to substantially broaden the distribution function. This process leads to an increase of plasma energy and is accompanied by a decrease of the field energy (Figure~\ref{fig:plasen}, dotted red). The increased absorptivity, in conjuction with the increased reflectivity, causes the transmissivity to decrease. The process saturates by \mbox{$t\simeq3$ ps} when the broadening of $f(\mathbf{v})$ ceases. 
After saturation, both the reflectivity and the absorptivity drop. The transmitted laser then fills up the simulation box again and the energy plateaus.

In comparison, in the SF1 scenario where a resonant F wave scatters the laser in the forward direction, the reflectivity is little affected. However, the transmissivity decreases drastically near \mbox{$B_0=15$ kT}, as shown in Figure~\ref{fig:absorption}. This is due to a strong increase of the absorptivity, the peak of which reaches $A\simeq 0.8$, indicating that more than $60\%$ of the incident energy is absorbed. 
By the end of our simulation at \mbox{$t=7$ ps}, this absorption mechanism has not saturated. The distribution function continues to broaden and the plasma energy continues to increase (Figure~\ref{fig:plasen}, large dashed black).
It is worth emphasizing that the absorption is due to forward scattering. In other words, for experiments that only monitor laser backscattering and use it as an indicator of plasma wave activities, the strong interactions via the F wave are invisible, and the resultant high electron temperature may appear surprising.

Finally, in the SF2 scenario, the backscattered laser undergoes secondary forward scattering via the F wave. Since the backscattered light is consumed by the fast-growing secondary instability, the reflectivity does not change much near \mbox{$B_0=10$ kT}. Nevertheless, the transmissivity decreases due to an increase of the absorptivity when the distribution function broadens. 
Since this is a secondary process, the absorptivity remains low at early time when the plasma energy only increases gradually. However, at later times, when the plasma waves fully develop and the energy transfer channels become more efficient, the absorptivity increases and the plasma energy rises rapidly. The increase of $U_P$ is correlated with a decrease of $U_F$ in Figure~\ref{fig:plasen} (dashed purple).
This absorption mechanism has not yet saturated by the end of our simulation.

\section{\label{sec:conclusion}Discussion}
We have shown that at resonant background magnetic field strengths, laser-to-plasma energy transfer is significantly increased via excitation of electron-cyclotron-like F waves. Large-amplitude F waves efficiently transfer energy to plasma electrons and lead to rapid heating, especially along the direction of the magnetic field. 
Analogous wave heating mechanisms for ions interacting with Alfv\'en-cyclotron waves have been investigated in the context of solar corona, \cite{Araneda2008Proton} through which the Sun's corona could be heated to temperatures many times beyond that of its surface. \cite{walsh2003}

Although the enhanced absorption relies on two separate processes, namely, the excitation of plasma waves and their subsequent collisionless damping, the undamped growth rate of plasma waves (Figure~\ref{fig:mvals}) turns out to be a good indicator of the overall absorptivity (Figure~\ref{fig:absorption}).
This is especially remarkable because the warm-fluid theory underlying $M$ has a somewhat different wave dispersion relation and likely slightly different growth rates than kinetic simulations.
A systematic comparison between PIC simulations and the warm-fluid theory shows excellent agreement up to $\sim10^2$ eV temperature. \cite{shi2022benchmarking}
The absorption spectrum shown in Figure~\ref{fig:absorption} have additional subpeaks and time-dependencies that we have not elaborated. Overall, the cluster of peaks, each due to a distinct physical mechanism, occur when the laser frequency is about twice the F-wave frequency, and may be collectively referred to as two-magnon decays. 
As the plasma density ramps up, two-magnon decays can occur at places far away from the critical density region, unlike two-plasmon decays that only occur near the quarter-critical region.

Our findings may shed light on recent magnetized implosion experiments at the National Ignition Facility, where temperatures hotter than expected are observed. \cite{moody2021,strozzi2021}  During implosions, the initial seed magnetic fields of tens of tesla undergo flux compression and potentially reach kilo-tesla level. The field strength may pass through resonances where laser energy absorption is strongly enhanced via mechanisms described in this paper.
Moreover, with subsequent collisions with highly charged ions, the resultant hot electrons can boost the radiation of x-rays. Although the requisite magnetic fields are large, if they become available, the enhanced absorption via two-magnon decays may substantially increase the brightness and temperature of laser-driven x-ray sources.

\begin{acknowledgments}
This work was performed under the auspices of the U.S. Department of Energy by Lawrence Livermore National Laboratory under Contract No. DE-AC52-07NA27344 and was supported by the LDRD program under tracking numbers 20-ERD-057 and 19-ERD-038.
\end{acknowledgments}

\section*{Data and code availability}
The data that support the findings of this study are openly available in zenodo at \url{https://doi.org/10.5281/zenodo.6839968}. The underlying simulation data are generated using version 4.17.10 of the EPOCH code at \url{https://github.com/Warwick-Plasma/epoch}. 
The analytical results are produced using version 2.2.0 of the Three-Wave-MATLAB code at \url{https://gitlab.com/seanYuanSHI/three-wave-matlab}.

\section*{Bibliography}
\providecommand{\noopsort}[1]{}\providecommand{\singleletter}[1]{#1}%

\end{document}